\documentclass[aps,twocolumn,showpacs,floatfix]{revtex4}
\usepackage{graphicx,amsmath,amssymb,times}
\usepackage{txfonts}
\usepackage{psfrag}

\begin{document}
    \newcommand{\p}[1]{\textrm{#1}}
 \newcommand{\mb}[1]{\mathbf{#1}}
\newcommand{\phipp}{\big|\phi_{\mb{p}}^{(+)}\big>}
\newcommand{\phipav}{|\phi_\mb{p}^\p{av}\rangle}
\newcommand{\pp}[1]{\big|\psi_{p}(#1)\big>}
\newcommand{\drdy}[1]{\sqrt{-R'(#1)}}
\newcommand{\rbf}{$^{85}$Rb}
\newcommand{\rbs}{$^{87}$Rb}
\newcommand{\kf}{$^{40}$K}
\newcommand{\na}{${^{23}}$Na}
\newcommand{\cs}{$^{133}$Cs}
\newcommand{\muK}{\:\mu\textrm{K}}
\newcommand{\phibav}{$|\phi_{\p{b}}^{\p{av}}\big>$}
\newcommand{\reffig}[1]{\mbox{Fig.~\ref{#1}}}
\newcommand{\refeq}[1]{\mbox{Eq.~(\ref{#1})}}
\newcommand{\rhopt}{\mbox{$\rho(p,t)$}\:}
\newcommand{\cites}[1]{\mbox{\cite{#1}}}

  \title{Association of molecules using a resonantly modulated magnetic field}

  \author{Thomas M. Hanna}
  \affiliation{Clarendon Laboratory, Department of Physics,
    University of Oxford, Parks Road, Oxford, OX1 3PU, UK}  
  \author{Thorsten K\"{o}hler}
  \affiliation{Clarendon Laboratory, Department of Physics,
    University of Oxford, Parks Road, Oxford, OX1 3PU, UK}
  \author{Keith Burnett}
  \affiliation{Clarendon Laboratory, Department of Physics,
    University of Oxford, Parks Road, Oxford, OX1 3PU, UK}
    
  \begin{abstract}

We study the process of associating molecules from atomic gases using a magnetic field modulation that is resonant with the molecular binding energy.
We show that maximal conversion is obtained by optimising the amplitude and frequency of the modulation for the particular temperature and density of the gas.
For small modulation amplitudes, resonant coupling of an unbound atom pair to a molecule occurs at a modulation frequency corresponding to the sum of the molecular binding energy and the relative kinetic energy of the atom pair. 
An atom pair with an off-resonant energy has a probability of association which oscillates with a frequency and time-varying amplitude which are primarily dependent on its detuning.
Increasing the amplitude of the modulation tends to result in less energetic atom pairs being resonantly coupled to the molecular state, and also alters the dynamics of the transfer from continuum states with off-resonant energies.
This leads to maxima and minima in the total conversion from the gas as a function of the modulation amplitude.
Increasing the temperature of the gas leads to an increase in the modulation frequency providing the best fit to the thermal distribution, and weakens the resonant frequency dependence of the conversion.
Mean-field effects can alter the optimal modulation frequency and lead to the excitation of higher modes. 
Our simulations predict that resonant association can be effective for binding energies of order $h\,\times 1$\,MHz.

    \end{abstract}

  \date{\today}
\maketitle

\section{Introduction}

Cold diatomic molecules are often produced from atomic gases by varying a magnetic field around a zero-energy resonance~\cite{donley02, regal03, review} or by photo\-association~\cite{doyle}. 
Generally applicable implementations of the former technique are linear ramps of the magnetic field across the resonance~\cite{regal03, str03, cub03, xu03, herbig03, gre03, duerr04prl, muk04, duerr04pra, volz05, thompson05b}, and fast switches to fields close to the resonance~\cites{donley02}. 
In addition, the long lifetime of cold $^6$Li$_2$ dimers allows them to be created by holding the magnetic field close to resonance, causing thermalisation of the atomic into a molecular gas~\cite{jochim03}.
In recent experiments, dimers of \rbf$_2$~\cite{thompson05} and \mbox{\rbf-\rbs~\cite{papp06}} were associated from cold atomic gases by applying a magnetic field modulation resonant with the molecular binding energy. 
This technique eliminates the need for the magnetic field to spend time in the near-resonant, strongly interacting region.
It therefore reduces the unwanted effect of heating~\cites{thompson05, hodby05} during the production of molecules. 
The narrow Fourier spectrum of the pulse accurately targets the molecular state, minimising the coupling to deeper bound states and highly energetic continuum states.
In a direct comparison to a linear ramp using the same apparatus, Thompson \textit{et~al.} reported more efficient conversion using resonant association~\cite{thompson05}. 
This technique has also been used as an accurate probe of molecular binding energy~\cites{thompson05, papp06}. 
In addition, radio-frequency pulses have been used to associate \mbox{$^{40}$K-\rbs} dimers~\cite{ospelkaus06}.

In this paper we study the resonant association of molecules from thermal and condensed gases. 
Our approach precisely accounts for the continuum of states in a gas. 
The transition amplitude from a pair of unbound atoms to the bound molecular state depends on the relative kinetic energy of the atom pair.
A resonant continuum energy exists, at which the transition amplitude to the molecular state increases linearly with time. 
At small modulation amplitudes, the resonant continuum energy is given by the difference between the energy corresponding to the modulation frequency and the molecular binding energy.
The distribution of atoms in different continuum states, all contributing to the molecular production, gives the total conversion a dependence on temperature. 
The width of the thermal distribution increases with the temperature of the gas, weakening the resonant behaviour of the molecular production.
The continuum distinguishes the current case from the association of atom pairs held in optical lattices~\cite{bertelsen06}, where the resonant modulation couples the discrete ground state of the tightly confining potential to the molecular bound state.

We find damped oscillations in the number of molecules produced in the short-time limit, as observed in Ref.~\cite{thompson05}. 
The damping is caused by the dephasing of the transition amplitudes from states across the continuum.
After the damping out of the initial oscillations, the conversion increases at a rate which displays resonant dependence on the modulation frequency.
Maximal conversion is achieved when the frequency and amplitude of the magnetic field modulation are together optimised for the temperature and density of the gas. 
The modulation amplitude required depends on the sensitivity of the molecular state to the magnetic field.
As the modulation amplitude increases states of lower continuum energy are resonantly coupled, until the zero-energy continuum state is reached.
Beyond this point all continuum states are coupled in a non-resonant manner.
There remain some modulation amplitudes where, for momenta close to the peak of the thermal distribution, the transition amplitude is large enough to lead to a revival in conversion efficiency.
Our calculations of molecular production for binding energies ranging from \mbox{$h \times 5\:\p{kHz}$} to \mbox{$h \times 1\:\p{MHz}$} predict that resonant association can be effective over this range.
We also examine short pulses in pure condensates, where the mean-field shift and the excitation of higher modes alter the dynamics. 
In condensates, the damping of oscillations in conversion due to the dephasing of the transition amplitudes from different continuum states is suppressed.

In Sec.~\ref{sec:numerics} we introduce the magnetic field profile used for resonant association, and also set up the notation used in our calculations. 
We then discuss the dynamics of the transition amplitude for a pair of atoms to a molecule, and the dependence this has on the continuum, in Sec.~\ref{sec:continuum}. 
In Sec.~\ref{sec:dependence} we examine in turn the effects of altering the duration, frequency and amplitude of the modulation on the efficiency of the molecular production. 
We also discuss the dependence of the conversion efficiency on the temperature and density of the atomic gas. 
In each section we discuss the results of Thompson \textit{et al.}~\cite{thompson05}, which formed the original motivation for our studies, and then consider resonant association under a broader range of conditions. 
We conclude in Sec.~\ref{sec:conclusion}. 

\section{Magnetic field sequence}
\label{sec:numerics}

In this section we introduce the magnetic field sequence used in resonant association, as well as the notation used in this paper.
Our calculations of molecular production from thermal gases use a two-channel approach~\cite{child74, moerdijk95, drummond98, timmermans99, mies00}. In the implementation of Ref.~\cites{goral04}, the two-channel, two-body Hamiltonian for the case of a time-varying magnetic field $B(t)$ is given by 
\begin{align}
H_\p{2B}(B(t))&=|\p{bg}\rangle H_\p{bg}\langle\p{bg}|
+W|\p{bg}\rangle\langle\p{cl}|\nonumber\\
&+|\p{cl}\rangle\langle\p{bg}|W
+|\p{cl}\rangle H_\p{cl}(B(t))
\langle\p{cl}|\, .
\label{eq:h2b}
\end{align}
Here, $W$ is the interchannel coupling between the entrance and closed channel spin configurations which are indicated by `bg' and `cl', respectively. Choosing the zero of energy to coincide with the dissociation threshold of the entrance channel makes the closed-channel Hamiltonian contain all of the magnetic field dependence of $H_\p{2B}(B(t))$. We make the single-resonance approximation~\cite{goral04}, neglecting all closed-channel states which are far detuned from $E = 0$. The single closed-channel state retained is referred to as the resonance level $\left|\phi_\p{res}\right>$, and is degenerate with the entrance-channel dissociation threshold at the field strength $B_{\p{res}}$, i.e. $E_{\p{res}}(B_{\p{res}})~=~0$. The closed-channel Hamiltonian is then given by
\begin{eqnarray}
H_{\p{cl}}(B(t)) & = & \left|\phi_{\p{res}}\right>\big[E_{\p{res}}^{\p{av}} + E_{\p{res}}^{\p{mod}}(t)\big]\left<\phi_{\p{res}}\right| .
\label{eq:eres}
\end{eqnarray}
Here $E_{\p{res}}^{\p{av}} = \frac{\partial E_{\p{res}}}{\partial B}(B_{\p{av}} - B_{\p{res}})$, $E_{\p{res}}^{\p{mod}}(t) = \frac{\partial E_{\p{res}}}{\partial B}[B(t) - B_{\p{av}}]$, $B_\p{av}$ is the average magnetic field during the pulse, and $\frac{\partial E_{\p{res}}}{\partial B}$ is the difference in magnetic moment between the entrance and closed channels. The measurable location $B_0$ of the singularity in the scattering length $a$ is shifted from $B_\p{res}$ by the interchannel coupling.
\begin{figure}[htbp]
	\centering
		\includegraphics[width=\columnwidth, clip]{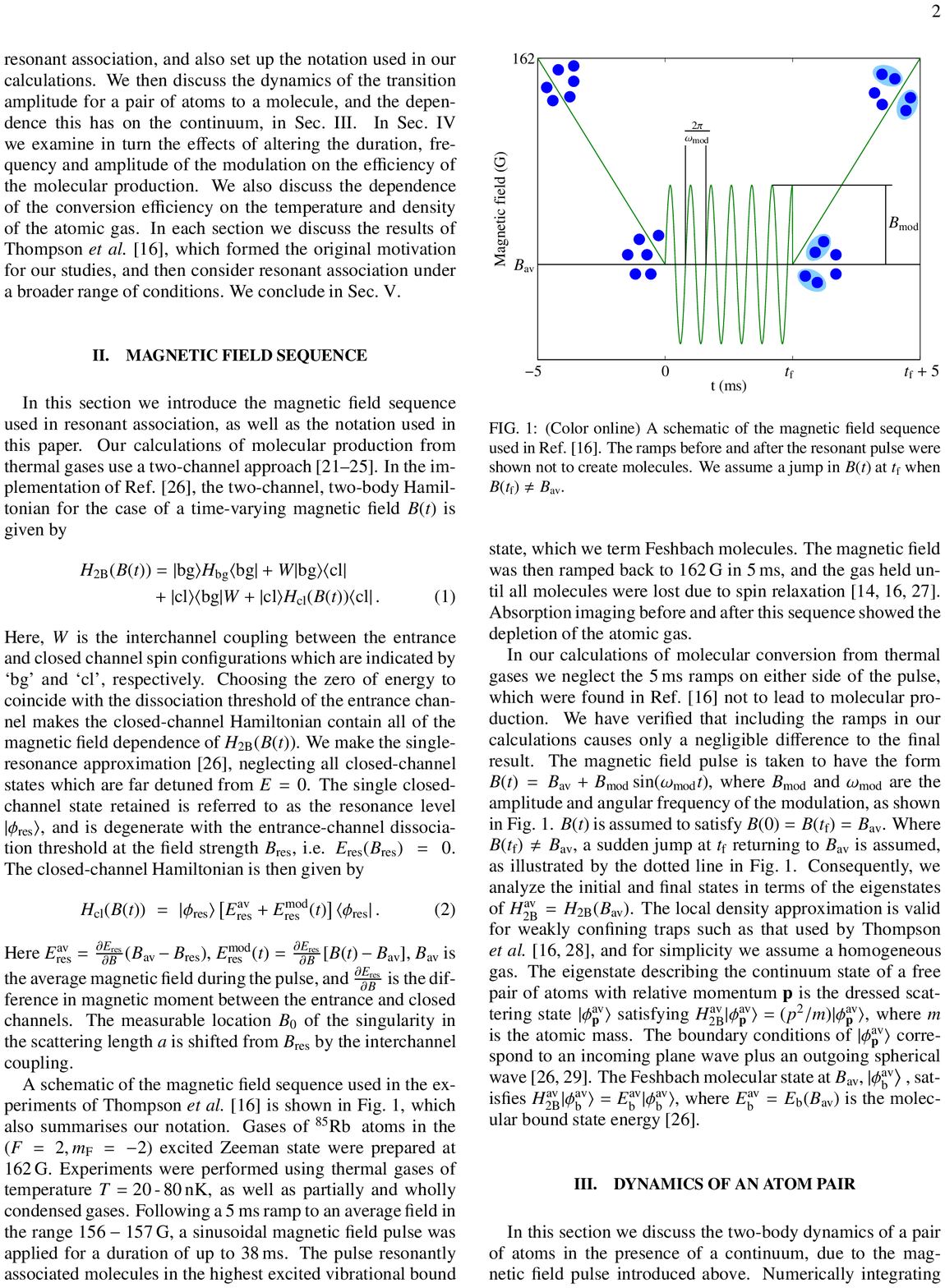}
	\caption{(Color online) A schematic of the magnetic field sequence used in Ref.~\cite{thompson05}. The ramps before and after the resonant pulse were shown not to create molecules. We assume a jump in $B(t)$ at $t_{\p{f}}$ when $B(t_{\p{f}}) \neq B_{\p{av}}$.}
	\label{fig:pulse}
\end{figure}

A schematic of the magnetic field sequence used in the experiments of Thompson \textit{et al.}~\cite{thompson05} is shown in \reffig{fig:pulse}, which also summarises our notation. Gases of \rbf\: atoms in the $(F = 2, m_\p{F} = -2)$ excited Zeeman state were prepared at 162\,G. Experiments were performed using thermal gases of temperature \mbox{$T = 20$\,-\,80\,nK}, as well as partially and wholly condensed gases. Following a \mbox{5\,ms} ramp to an average field in the range \mbox{$156 - 157$\,G}, a sinusoidal magnetic field pulse was applied for a duration of up to \mbox{38\,ms}. The pulse resonantly associated molecules in the highest excited vibrational bound state, which we term Fesh\-bach molecules. The magnetic field was then ramped back to \mbox{162\,G} in \mbox{5\,ms}, and the gas held until all molecules were lost due to spin relaxation~\cites{thompson05, thompson05b, koehler05}. Absorption imaging before and after this sequence showed the depletion of the atomic gas.

In our calculations of molecular conversion from thermal gases we neglect the \mbox{5\,ms} ramps on either side of the pulse, which were found in Ref.~\cite{thompson05} not to lead to molecular production. We have verified that including the ramps in our calculations causes only a negligible difference to the final result. The magnetic field pulse is taken to have the form $B(t) = B_\p{av} + B_\p{mod}\sin(\omega_\p{mod}t)$, where $B_\p{mod}$ and $\omega_\p{mod}$ are the amplitude and angular frequency of the modulation, as shown in \reffig{fig:pulse}. $B(t)$ is assumed to satisfy $B(0) = B(t_{\p{f}}) = B_{\p{av}}$. Where $B(t_{\p{f}}) \neq B_{\p{av}}$, a sudden jump at $t_{\p{f}}$ returning to $B_{\p{av}}$ is assumed, as illustrated in \reffig{fig:pulse}. Consequently, we analyze the initial and final states in terms of the eigenstates of $H_\p{2B}^\p{av} = H_\p{2B}(B_\p{av})$. The local density approximation is valid for weakly confining traps such as that used by Thompson \textit{et al.}~\cites{thompson05, cornish00}, and for simplicity we assume a homogeneous gas. The eigenstate describing the continuum state of a free pair of atoms with relative momentum $\mb{p}$ is the dressed scattering state $\phipav$ satisfying \mbox{$H_\p{2B}^\p{av}\phipav = (p^{2}/m)\phipav$}, where $m$ is the atomic mass. The boundary conditions of $\phipav$ correspond to an incoming plane wave plus an outgoing spherical wave~\cite{taylor72, goral04}. The Fesh\-bach molecular state at $B_\p{av}$, \phibav\:, satisfies \mbox{$H_\p{2B}^\p{av}|\phi_\p{b}^\p{av}\rangle = E_\p{b}^\p{av}|\phi_\p{b}^\p{av}\rangle$}, where $E_\p{b}^\p{av} = E_\p{b}(B_\p{av})$ is the molecular bound state energy~\cite{goral04}.


\section{Dynamics of an atom pair}
\label{sec:continuum}

In this section we discuss the two-body dynamics of a pair of atoms in the presence of a continuum, due to the magnetic field pulse introduced above.
Numerically integrating the Schr\"{o}dinger equation associated with the Hamiltonian of \refeq{eq:h2b} for a particular pulse gives the two-body time evolution operator $U_{\p{2B}}(t_{\p{f}},0)$, linked to $H_{\p{2B}}(B(t))$ by
\begin{align}
  i\hbar\frac{\partial}{\partial t}U_{\p{2B}}(t,t')
  =H_{\p{2B}}(B(t))U_{\p{2B}}(t,t') \, .
  \label{eq:u2b}
\end{align}
After the pulse, the wavefunction of a pair of atoms that were initially in the state $\phipav$ has an overlap with the Fesh\-bach molecular state given by the transition amplitude
\begin{equation}
T(p, t_{\p{f}}) = \langle\phi_\p{b}^\p{av}|U_{\p{2B}}(t_{\p{f}},0)\phipav.
\label{eq:T}
\end{equation}
This in turn gives the probability density for a transition between a free pair of atoms of relative momentum $\mb{p}$ and a Fesh\-bach molecule to be
\begin{equation}
\rho(p,t_{\p{f}}) = \left|T(p,t_{\p{f}})\right|^{2} \, .
\label{eq:rho}
\end{equation}
In this paper we consider resonances where a spherically symmetric resonance level is coupled to the entrance channel by spin exchange~\cite{review}. Consequently, the transition amplitude and probability density depend only on the modulus of the momentum. The transition probability density gives the dynamics of only one state in the continuum. This is to be distinguished from the conversion itself, which includes the contributions of all the continuum states.

In the limit of short times and a small modulation amplitude, $U_{\p{2B}}(t_{\p{f}},0)$ may be approximated using time-dependent perturbation theory. Treating the oscillating component of the Hamiltonian of \refeq{eq:eres} as a perturbation to $H_\p{2B}^\p{av}$ gives the first order approximation to the two-body evolution operator: 
\begin{align}
U_\p{2B}^{(1)}(t_\p{f},0) & \approx U_{\p{2B}}^\p{av}(t_\p{f}) \nonumber \\ 
& +  \frac{1}{i\hbar}\int_0^{t_\p{f}} dt \: U_\p{2B}^\p{av}(t_\p{f} - t)|\phi_\p{res}\rangle E_\p{res}^\p{mod}(t)\langle \phi_\p{res} | U_{\p{2B}}^{\p{av}}(t) \, .
\label{eq:U2B1}
\end{align}
Here, the two-body evolution operator at $B_\p{av}$ is given by $U_{\p{2B}}^{\p{av}}(t) = \exp(-i H_\p{2B}^\p{av}t/\hbar)$. Projecting the estimate of \refeq{eq:U2B1} onto $\langle\phi_{\p{b}}^{\p{av}} |$ on the left and $|\phi_{\mb{p}}^\p{av}\rangle$ on the right gives an approximation to the transition amplitude of \refeq{eq:T}: 
\begin{align}
  T^{(1)}(p,t_\p{f}) &= -\frac{1}{2\hbar}\frac{\partial E_{\p{res}}}{\partial B}B_{\p{mod}}e^{-i E_\p{b}^\p{av} t_\p{f}/\hbar} C(p) \, \nonumber \\
  &\times\bigg[
  e^{i\omega_+t_\p{f}}
  \frac{\sin(\omega_+t_\p{f})}{\omega_+} - e^{i\omega_-t_\p{f}}
  \frac{\sin(\omega_-t_\p{f})}{\omega_-}
  \bigg]\, .
  \label{eq:T1}
\end{align}
Here
\begin{align}
  \omega_\pm=
  \left(
  E_\p{b}^\p{av}\pm\hbar\omega_\p{mod}-p^2/m \right)/(2\hbar) \,,
  \label{eq:wpm}
\end{align}
and $C(p)= \big<\phi_{\p{b}}^{\p{av}}\big|\phi_{\p{res}}\big>\big<\phi_{\p{res}}\phipav$ is the product of the overlaps of the resonance state with the bound and scattering states at $B_{\p{av}}$. 
Since $E_\p{b}^\p{av}$, $\hbar \omega_{\p{mod}}$ and $p^{2}/m$ can all be of the same order of magnitude, it is not in general possible to make the rotating wave approximation and neglect the $\omega_-$ term in \refeq{eq:T1}. 

\begin{figure}[htbp]
	\centering
		\includegraphics[width=\columnwidth, clip]{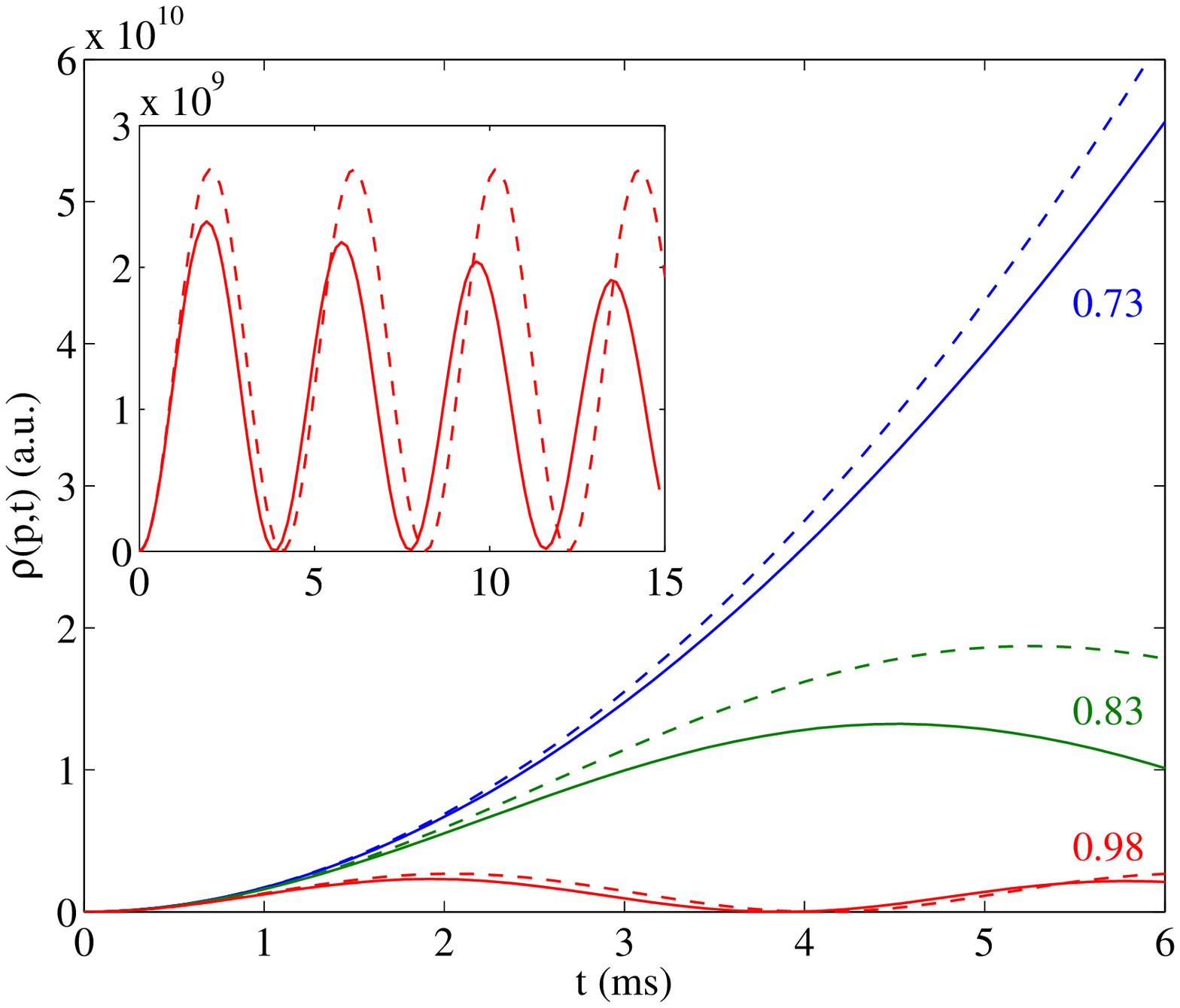}
	\caption{(Color online) Evolution of the transition probability density \rhopt in a \rbf\:gas for near-resonant continuum energies. The resonance condition of \refeq{eq:res} is fulfilled at \mbox{$p^{2}/m = h \times 0.73$\,kHz}. The solid lines show the numerical results, and the dashed lines show the perturbative estimate of \refeq{eq:T1}. The transition probability densities shown are labelled with the energy of the continuum state in $h \,\times\,$kHz. The inset shows the evolution of \rhopt for the \mbox{0.98\,kHz} continuum state at longer times. Here, \mbox{$B_{\p{av}} = 156.45$\,G}, \mbox{$B_{\p{mod}} = 0.065$\,G}, \mbox{$E_{\p{b}}^{\p{av}}/h = -5.77$\,kHz}, and \mbox{$\omega_{\p{mod}}/2\pi = 6.5$\,kHz.}}
	\label{fig:mat_res}
\end{figure}

As implied by the first-order estimate of \refeq{eq:T1}, the fastest growth in the transition probability density \rhopt for small $B_\p{mod}$ occurs for the resonant continuum energy $p_\p{res}^2/m$, which satisfies 
\begin{equation}
E_\p{b}^\p{av} + \hbar\omega_{\p{mod}} - \frac{p_{\p{res}}^{2}}{m} = 0 \, .
\label{eq:res}
\end{equation}
This corresponds to the sum of the relative kinetic energy of the atom pair and the molecular binding energy $|E_\p{b}^\p{av}|$ being exactly matched by the modulation frequency. As shown in \reffig{fig:mat_res}, the growth in the transition probability density \rhopt for \mbox{$p^{2}/(mh) = 0.73$\,kHz} is quadratic. This corresponds to the transition amplitude of \refeq{eq:T} having a linearly increasing amplitude, similar to a resonantly driven harmonic oscillator.

At sufficiently short times, \rhopt grows quadratically for states detuned from the resonant continuum energy. \mbox{Figure~\ref{fig:mat_res}} shows that \rhopt displays behaviour different to $\rho(p_{\p{res}},t)$ after a time of order \mbox{$\hbar/(|p^{2} - p_{\p{res}}^{2}|/m)$}. The inset of \reffig{fig:mat_res} shows the oscillatory nature of \rhopt, reproduced at short times by the analytic estimate of \refeq{eq:T1}. 
The numerical approach, accounting precisely for the continuum of states, yields an envelope in the oscillation amplitude as well as a frequency shift from the estimate of \refeq{eq:T1}.
The thermal gases measured in Ref.~\cite{thompson05} had temperatures in the range 20\,-\,80\,nK. The energy $k_\p{B} \times 50$\,nK corresponds to $h \times 1$\,kHz. Consequently, the phase difference between the continuum states spread over the thermal distribution becomes significant after times of the order of milliseconds.


\section{Experimental parameters affecting the conversion efficiency}
\label{sec:dependence}

In this section we study the variation of the molecular production with the duration, frequency and amplitude of the magnetic field modulation, and the temperature and density of a thermal or fully condensed gas. 
We refer to the fraction of atoms converted to molecules as the conversion efficiency. 
In the limit of small depletion of a thermal atomic gas, this is given by a weighted average of the transition probability density $\rho(p,t_{\p{f}})$ over a Maxwell distribution:
\begin{equation}
\frac{2N_{\p{mol}}}{N} = 2n(2\pi\hbar)^{3}\left(\frac{\beta}{\pi m}\right)^{3/2} \int d\mb{p} \exp\left(\frac{-\beta p^{2}}{m}\right) \rho(p,t_{\p{f}}) \, .
\label{eq:conv}
\end{equation}
Here, $\beta = 1/k_{\p{B}}T$, $n$ is the density of the atomic gas, $N$ is the initial number of atoms and $N_\p{mol}$ is the final number of molecules.
In this limit the conversion efficiency is proportional to the density of the atomic gas. 
We note that this approach does not lead to saturation of the conversion efficiency, which would require the inclusion of genuinely many-body effects. This requires the solution of non-Markovian Boltzmann-like equations, whose Markov limit has previously been used to study the special case of saturation of molecular production from magnetic field ramps~\cite{williams06}.

\subsection{Pulse duration}
\label{sec:time}

\subsubsection{Thermal gas}

\begin{figure}[htbp]
	\centering
		\includegraphics[width=\columnwidth, clip]{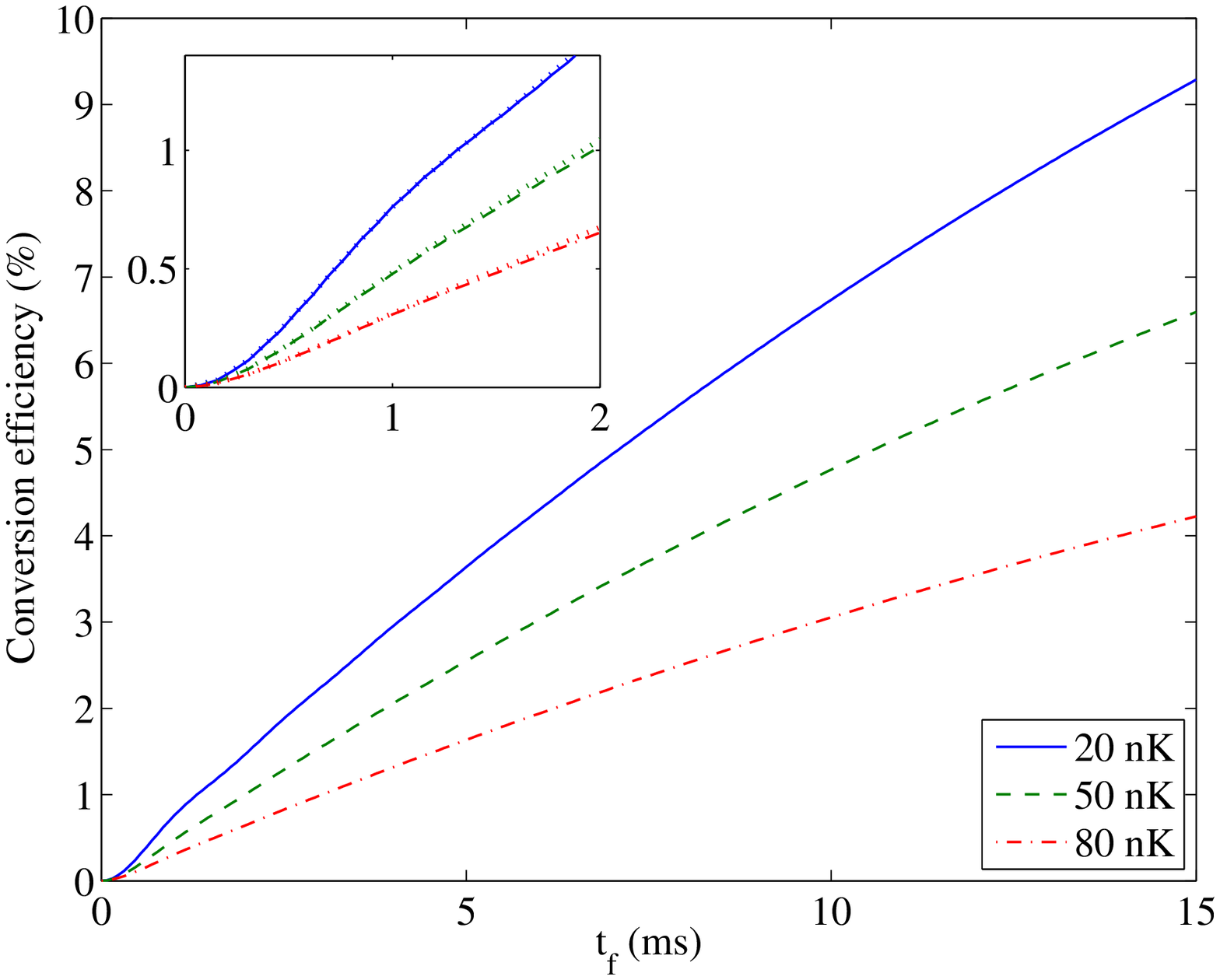}
	\caption{(Color online) Conversion efficiency from a thermal \rbf\: gas as a function of pulse duration, for  a density of \mbox{$n = 10^{11}$\,cm$^{-3}$}. All other parameters are the same as those of the data represented in \reffig{fig:mat_res}, which has been thermally averaged according to \refeq{eq:conv} to give the conversion efficiency. The inset shows damped oscillations in the conversion efficiency, visible for \mbox{$T = 20$\,nK} but washed out for 50\,nK by the dephasing of the transition amplitudes from different continuum states. The dotted lines show the results given by thermally averaging the perturbative estimate of \refeq{eq:T1}.
} 
	\label{fig:time}
\end{figure}
We first consider the conversion efficiency from a thermal gas.
The averaging of \refeq{eq:conv} gives a contribution from \rhopt for each $p$, weighted according to the thermal distribution. 
\mbox{Figure~\ref{fig:time}} shows the resulting conversion efficiency for gases of 20, 50 and \mbox{80\,nK} as a function of pulse duration. 
The resonance condition of \refeq{eq:res} is fulfilled at a continuum energy of \mbox{$h\, \times\, 0.73$\,kHz}, which corresponds to \mbox{37\,nK}. 
Of the gas temperatures quoted in Ref.~\cite{thompson05}, \mbox{20\,nK} gives the highest conversion efficiency because the most atom pairs have energies close to the resonant continuum energy. 

In the experiments of Ref.~\cite{thompson05}, damped oscillations in the conversion efficiency as a function of time were observed over the first few milliseconds. In our calculations, damped oscillations are visible over approximately \mbox{2\,ms} for a temperature of \mbox{20\,nK}. We have verified for several values of $B_{\p{av}}$ and $\omega_{\p{mod}}$ that the frequency of the damped oscillations, $f_\p{conv}$, for the thermal gas case is close to $p_{\p{res}}^{2}/(mh)$. This is the value of $\omega_{+}/2\pi$ for \mbox{$p = 0$} in \refeq{eq:wpm}, corresponding to the detuning of the zero momentum state from the resonant continuum energy. 
Increasing the temperature causes a negative shift in the frequency of the damped oscillations, together with faster damping. 
For 50 and \mbox{80\,nK} gases our calculations do not predict damped oscillations large enough to be observed. The main cause of damping is the variation in the oscillation frequency of \rhopt with $p$, as shown in \reffig{fig:mat_res} and discussed in Sec.~\ref{sec:continuum}. A wider thermal distribution corresponds to a wider spread in momentum of the atom pairs contributing to the conversion efficiency, and so the initial coherence in \rhopt across the distribution is destroyed more quickly. 

\subsubsection{Condensed gas}

The critical temperature reported in Ref.~\cite{thompson05} is \mbox{14\,nK}, with average densities of order \mbox{$10^{11}\,$cm$^{-3}$}. For a density of \mbox{$10^{11}\,$cm$^{-3}$}, which we use in our thermal gas calculations, and a magnetic field of \mbox{156.45\,G}, the dilute gas parameter $\sqrt{n a^{3}}$ is 0.02. For such gases, which are close to condensation or partially condensed, there will be a mean field shift in the frequency of the oscillations in conversion efficiency. We have analyzed this effect for the case of a pure, homogeneous condensate. For our studies of condensed gases we use the cumulant approach~\cites{koehler03pra, koehler03prl, koehler02pra}. 
In this approach, the atomic mean field $\Psi(t)$ is given by a non-Markovian, nonlinear Schr\"odinger equation~\cite{koehler03pra}:
\begin{align}
i\hbar \frac{\partial}{\partial t}\Psi(t) = H_\p{1B}\Psi(t)
- \Psi^{*}(t) \int_{0}^\infty d\tau \, \Psi^2(\tau) \frac{\partial}{\partial \tau} h(t, \tau) \, .
\label{nlse}
\end{align}
The coupling function $h(t, \tau)$ contains the exact two-body dynamics, and is given by
\begin{align}
h(t, \tau) = (2\pi\hbar)^3\big<0|V(t)U_{\p{2B}}(t,\tau)|0\big>\theta(t-\tau) \, .
\label{httau}
\end{align}
Here, the evolution operator $U_\p{2B}(t, \tau)$ is defined in \refeq{eq:u2b}, $|0\big>$ is the plane wave of zero momentum, $V(t)$ is the diatomic potential, and \mbox{$\theta(t - \tau)$} is the step function, yielding 1 for \mbox{$t > \tau$} and 0 otherwise. The molecular conversion is given by a molecular mean field,
\begin{align}
\Psi_\p{b}(t) = -\frac{1}{\sqrt{2}}\int_{0}^\infty d\tau \, \Psi^2(\tau)\frac{\partial}{\partial \tau} h_\p{b}(t, \tau) \, ,
\label{psib}
\end{align}
where the bound state coupling function is given by
\begin{align}
h_\p{b}(t, \tau) = (2\pi\hbar)^{3/2}\big<\phi_\p{b}|U_{\p{2B}}(t,\tau)|0\big>\theta(t-\tau) \, .
\label{hbttau}
\end{align}
The coupling functions of \mbox{Eqs.~\eqref{httau} and~\eqref{hbttau}} have been determined from the single-channel approach of Ref.~\cite{koehler03pra}. The magnetic fields used in this calculation are within the range for which single-channel approaches have been shown to be valid for the 155\,G resonance of \rbf~\cites{koehler04}.

\begin{figure}[htbp]
	\centering
		\includegraphics[width=\columnwidth, clip]{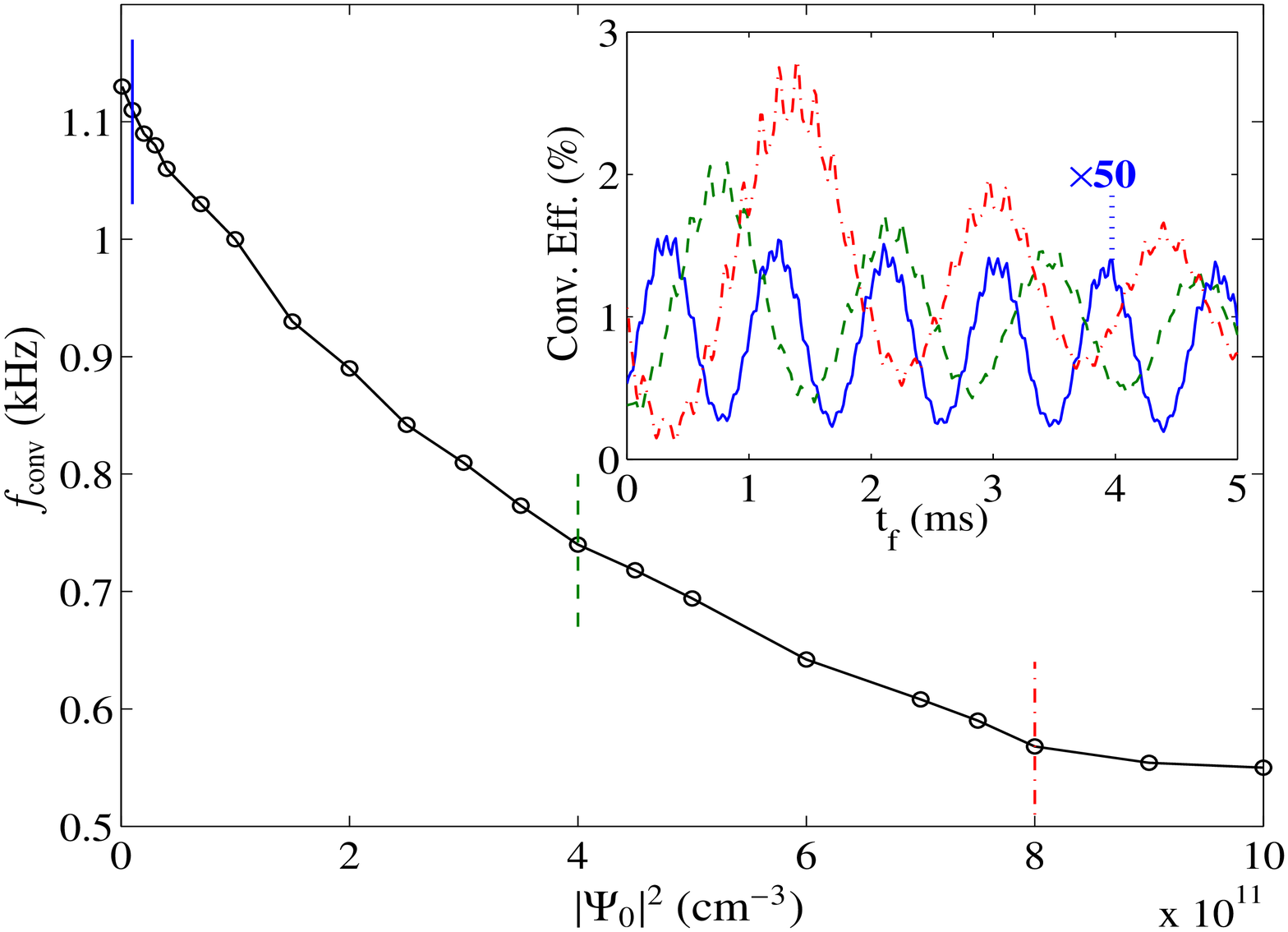}
	\caption{(Color online) Frequency of the oscillations in conversion efficiency at short times for a pure \rbf\: condensate as a function of the initial density. The mean-field shift lowers the frequency of the oscillations in conversion efficiency from that given by the two-body approach, which is recovered in the limit of low density. The oscillations are, however, much clearer than those in a thermal gas due to the suppression of the contributions of different continuum states to the molecular production. The inset shows the variation in time of the conversion efficiency for densities of $10^{10}$ ($\times 50$ for clarity), $4\times10^{11}$ and \mbox{$8 \times 10^{11}$\,cm$^{-3}$}. The \mbox{0.5\,ms} ramp from \mbox{B = 157.45\,G} to \mbox{$B_\p{av} = 156.45$\,G}, not shown here, gives a density-dependent initial phase to the oscillations. Here \mbox{$B_\p{mod} = 0.065$\,G}, \mbox{$E_\p{b}^\p{av}/h = -5.86$\,kHz} and \mbox{$\omega_\p{mod}/2\pi = 7$\,kHz}. Because a single-channel approach is used in the condensed gas case, the bound state energy is slightly different to that given by the two-channel, thermal gas calculations above.}
	\label{fig:condensate}
\end{figure}

The oscillation frequency of the conversion efficiency at short times, $f_\p{conv}$, has a mean-field shift, as shown in \reffig{fig:condensate}. 
In the low-density limit, the value of $f_\p{conv}$ expected from the two-body picture is recovered. 
The oscillations in conversion efficiency are clearer and have weaker damping than those in thermal gases, as shown in the inset of \reffig{fig:condensate}.
This is due to suppression of the dephasing between the transition amplitudes from different continuum states. 
The main cause of damping in this case is the decay of the condensate and molecular populations into the continuum. 
In these calculations we have included a \mbox{0.5\,ms} ramp from \mbox{$B_\p{av} + 1$\,G} to $B_\p{av}$, in analogy to the ramp shown in \reffig{fig:pulse}. 
Neglecting the ramp and simulating only the pulse corresponds to instantly turning on the interactions at the beginning of the pulse.
For the parameters used here, this results in strong excitation of higher modes.
The ramp reduces but does not completely eliminate the excitations, which are visible in the inset of \reffig{fig:condensate} as high frequency oscillations whose amplitude increases with condensate density.
We have extracted the frequency of the damped oscillations using the fit procedure of Claussen \textit{et al.}~\cite{claussen03}, which includes exponential damping of the oscillations and a linear decay. 
Strong decay of the condensate into the continuum at higher densities makes the fit less reliable and meaningful, and we have therefore limited the analysis using this technique to densities below \mbox{$10^{12}\,$cm$^{-3}$}. 

\subsection{Modulation frequency}
\label{sec:freq}

\begin{figure}[htbp]
	\centering
		\includegraphics[width=\columnwidth, clip]{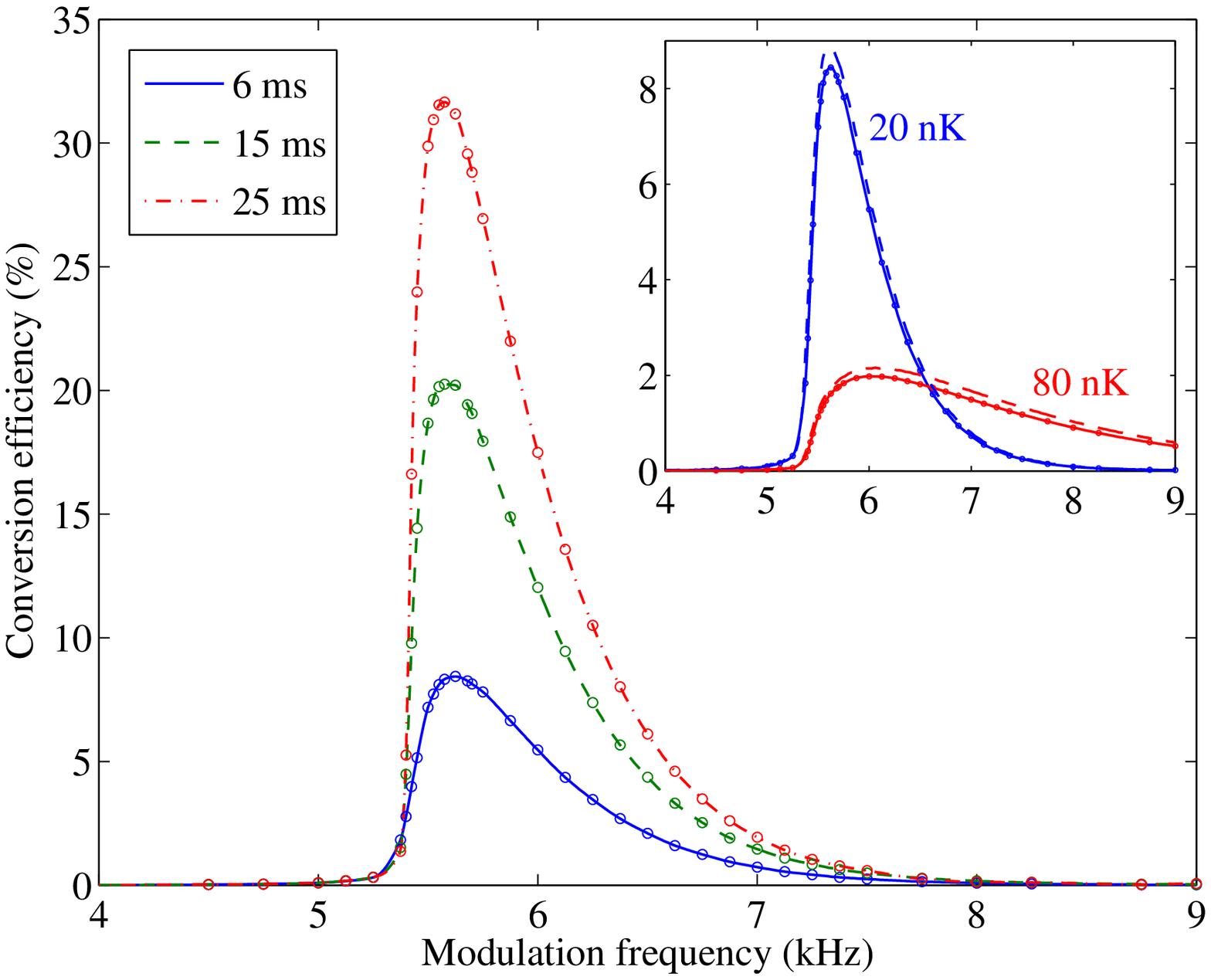}
	\caption{(Color online) Resonance curve of conversion efficiency vs modulation frequency for different pulse durations, for a thermal \rbf\: gas. Here \mbox{$B_{\p{av}} = 156.41$\,G}, \mbox{$E_\p{b}^\p{av}/h = -5.39$\,kHz}, \mbox{$B_{\p{mod}}= 0.065$\,G}, \mbox{$T = 20$\,nK} and \mbox{$n = 10^{11}$\,cm$^{-3}$}. Inset: Conversion efficiency after a \mbox{6\,ms} pulse with \mbox{$T = 20$\,nK} and \mbox{$T = 80$\,nK,} showing its weaker dependence on modulation frequency at higher temperatures. The dashed curves are thermal averages of the perturbative estimate of \refeq{eq:T1} for \mbox{6\,ms} pulses.}
	\label{fig:freq}
\end{figure}

Resonant behaviour was observed in Ref.~\cite{thompson05} in the strong variation of the conversion efficiency with modulation frequency, which is reproduced by our calculations. 
The conversion efficiency from a thermal gas due to a pulse of fixed duration and varying frequency is shown in \reffig{fig:freq}. 
For the resonance curve representing \mbox{6\,ms} pulses, the full-width at half-maximum is \mbox{0.75\,kHz}. From the Lorentzian fit to the \mbox{6\,ms} pulse in \mbox{Fig. 1} of Ref.~\cite{thompson05} we extract \mbox{0.9\,kHz}. 
At longer times many-body effects may lead to the production of molecules by thermalisation. This could lead, for example, to the production of molecules for modulation frequencies smaller than $-E_\p{b}^\p{av}/h$, and so increase the width of the resonance curve. 
The inset of \reffig{fig:freq} shows estimates of the conversion efficiency using the perturbative estimate of the transition amplitude in \refeq{eq:T1}. 
The agreement with the numerical result is significantly better than that of the transition probability density, shown in \reffig{fig:mat_res}, due to the effect of thermally averaging over all of the continuum states.  

The maximum of a thermal distribution is at a higher energy in a warmer gas, and so the optimal modulation frequency increases with temperature. 
However, the dependence of the conversion efficiency on modulation frequency weakens at higher temperatures, as shown in the inset of \reffig{fig:freq}. 
This is caused by the changes in the thermal distribution of the gas, which has a decreasing maximum and an increasing width as the temperature rises. 
The decreasing maximum of the distribution leads to less being gained by optimising the modulation frequency $\omega_\p{mod}/2\pi$. 
Conversely, the increasing width means that a wider range of $\omega_\p{mod}$ have a significant population of atoms close to the resonant continuum energy $p_\p{res}^2/m = E_\p{b}^\p{av} + \hbar \omega_\p{mod}$. 
In general, the stronger resonant behaviour in colder gases allows more efficient conversion. 

\begin{figure}[htbp]
	\centering
		\includegraphics[width=\columnwidth, clip]{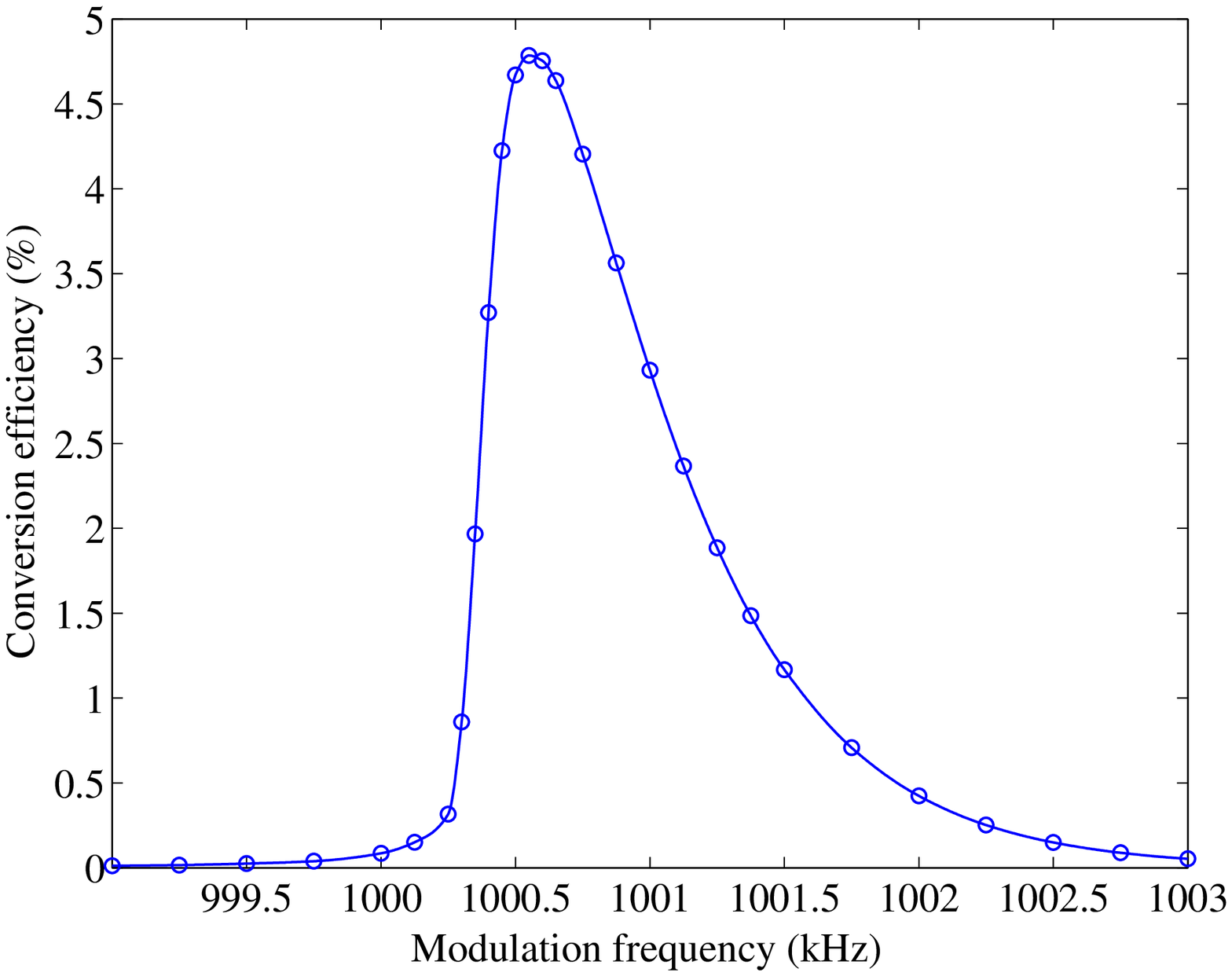}
	\caption{(Color online) Conversion efficiency vs modulation frequency for a thermal $^{133}$Cs gas, from a 6\,ms pulse. The gas density and modulation amplitude are identical to the \rbf\: curve shown in \reffig{fig:freq}. Here \mbox{$T = 20$\,nK}, \mbox{$n = 1 \times 10^{11}\,$cm$^{-3}$}, \mbox{$B_\p{av} = 21.37$\,G}, \mbox{$B_\p{mod} = 0.065$\,G}, and \mbox{$E_\p{b}^\p{av}/h = -1$\,MHz}.}
	\label{fig:csfreq}
\end{figure}
We have studied the conversion efficiency vs frequency for molecular binding energies of $h\times100$\,kHz, and found that it leads to a resonance curve of similar width and maximum to that for the binding energies examined above. 
\rbf$_2$, though, is unstable with respect to inelastic spin relaxation~\cites{thompson05b, koehler05}. 
We have therefore also performed the calculation for $^{133}$Cs atoms in the \mbox{($F = 3$, $m_\p{F} = 3$)} Zeeman ground state for a molecular bound state energy of \mbox{$-h \times 1$\,MHz}. 
A similar resonance curve is obtained, as shown in \reffig{fig:csfreq}. 
It is broader and has a lower maximum than the comparable calculations for 6\,ms pulses in \rbf, which had an identical modulation amplitude and gas density. 
Despite the deeper binding energy, the conversion efficiency grows at a similar rate. 
The evolution of the transition probability density for a continuum state depends primarily upon its detuning from the resonant continuum energy. 
Consequently, it is primarily the width of the thermal distribution, rather than the molecular bound state energy, that determines the order of magnitude of the pulse duration necessary for association.
Our calculations indicate that resonant association can be efficient for binding energies ranging from $h \times 5$\,kHz to $h \times 1$\,MHz. 
It is necessary, however, that the chosen molecular binding energy be sensitive to variations in the magnetic field.
If this is not the case, the magnetic field modulation has little or no effect on the diatomic level spectrum, and so significant transitions between the continuum states and the molecular bound state do not occur. 
Such a weak dependence on the magnetic field can occur due to an avoided crossing with another bound state, as occurs for $^{133}$Cs$_2$ at some binding energies~\cite{chin04}.
In some cases, it may be possible to compensate for this by using a larger modulation amplitude.

\subsection{Modulation amplitude}
\label{sec:bosc}

\begin{figure}[htbp]
	\centering
		\psfrag{LAB}{(a)}
		\includegraphics[width=\columnwidth, clip]{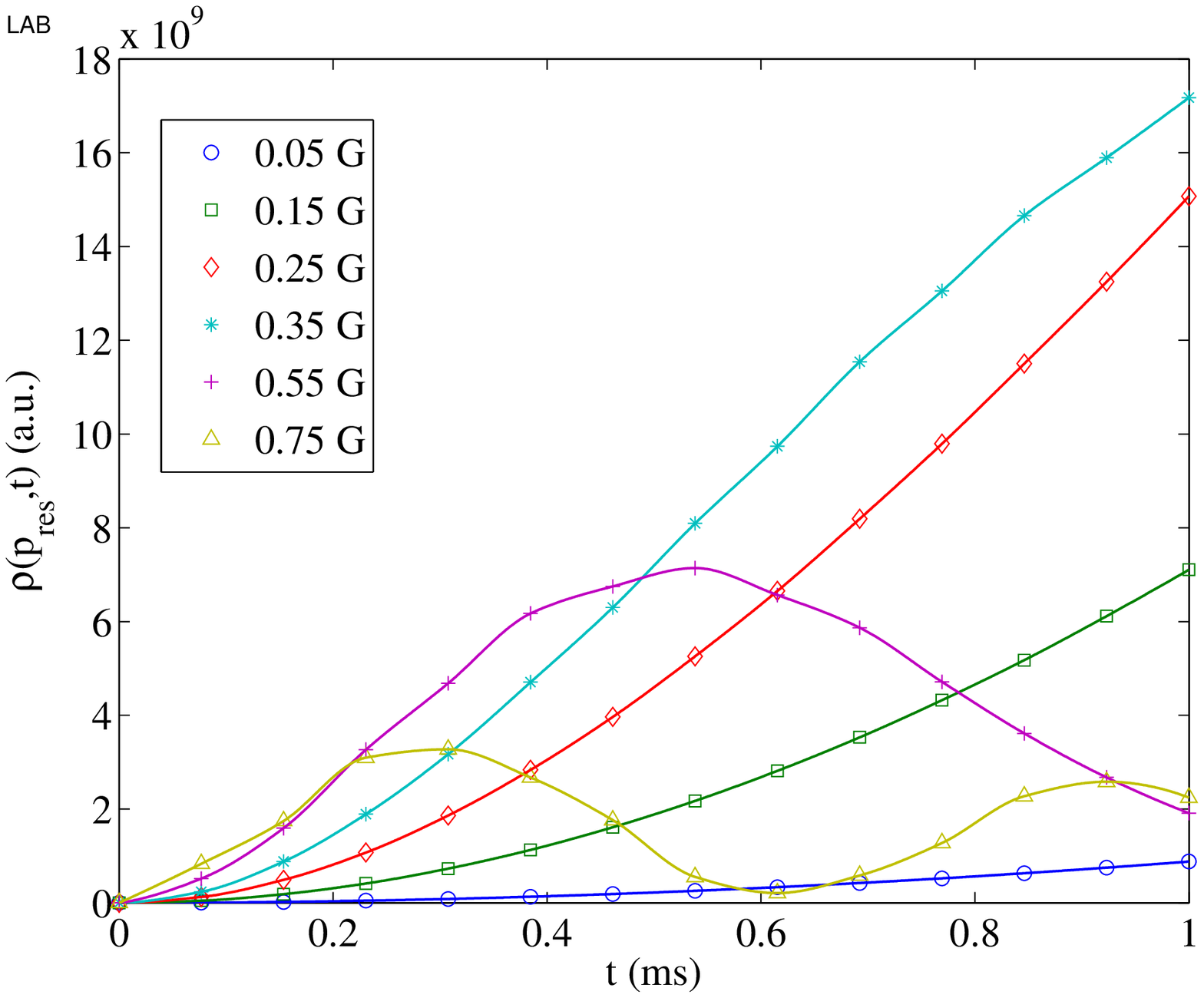}
		\psfrag{LAC}{(b)}
		\includegraphics[width=\columnwidth, clip]{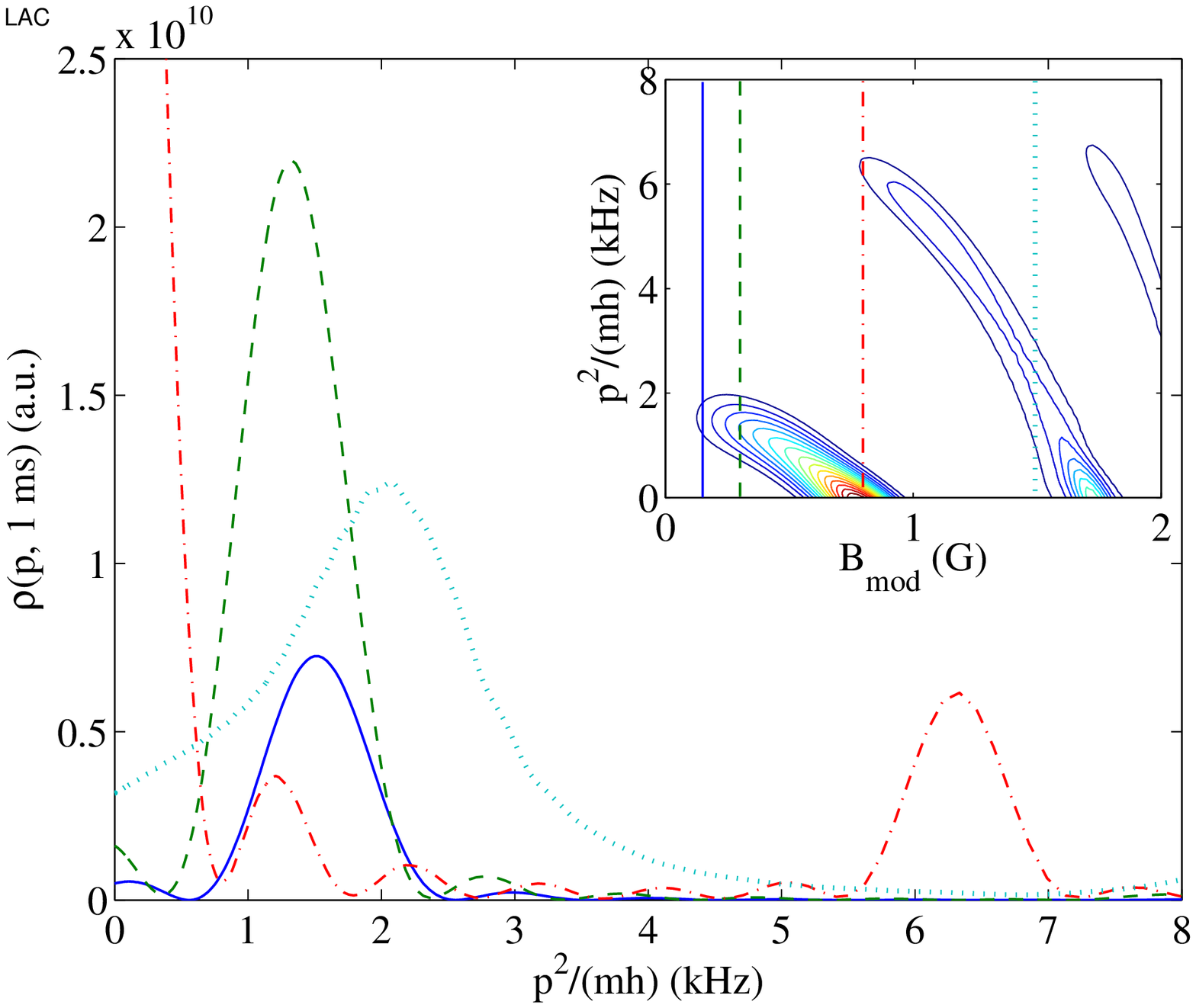}
		
	\caption{(Color online) The transition probability density \rhopt in a \rbf\: gas for different values of $B_{\p{mod}}$. Here \mbox{$B_{\p{av}} = 156.352$\,G}, \mbox{$\omega_{\p{mod}}/2\pi = 6.5$\,kHz}, and \mbox{$E_{\p{b}}^{\p{av}}/h = -4.88$\,kHz}. \mbox{(a) The} evolution of \rhopt for the continuum state of energy satisfying the resonance condition of \refeq{eq:res}. Quadratic growth ceases to be observed when the modulation amplitude becomes too great. (b) The transition probability density distribution \mbox{$\rho(p, 1\,\p{ms})$} for $B_\p{mod} = $ 0.15, 0.3, 0.8 and 1.5\,G, as indicated in the contour plot shown in the inset. For each \mbox{$B_{\p{mod}}<0.9$\,G}, the peak in energy of \mbox{$\rho(p, t)$} grows resonantly on a timescale of \mbox{1\,ms}. The continuum energy of the maximal \mbox{$\rho(p, 1\,\p{ms})$} is negatively shifted from $p_{\p{res}}^{2}/m$ with increasing $B_{\p{mod}}$. Two bands of revival in \rhopt can be seen in the inset, as well as in the distributions of \rhopt for $B_\p{mod} = 0.8$ and 1.5\,G.}
	\label{fig:mat_bosc}
\end{figure}

In the experiments of Thompson \textit{et al.}, increasing the modulation amplitude $B_{\p{mod}}$ with a fixed frequency and pulse duration gave a point of maximum conversion, and after reaching a minimum a partial revival was observed~\cite{thompsonpc}. 
Examining the transition probability density of \refeq{eq:rho} for different $B_\p{mod}$ shows that as $B_{\p{mod}}$ is increased, the resonant growth of $\rho(p_{\p{res}},t)$ is at first amplified, as shown in \reffig{fig:mat_bosc}a. The faster resonant growth is also reflected in the proportionality of the analytic estimate of $T(p, t)$ in \refeq{eq:T1} to $B_{\p{mod}}$. For \mbox{$B_{\p{mod}} = 0.35$\,G}, there is no longer resonant growth in \mbox{$\rho(p_{\p{res}},t)$} over a \mbox{1\,ms} pulse duration, although this modulation amplitude does maximize \mbox{$\rho(p_{\p{res}}, 1\,\p{ms})$}. The changing amplitude and position of the maximum as $B_\p{mod}$ varies alters the quality of the fit to the thermal distribution, and consequently the conversion efficiency.

\begin{figure}[htbp]
	\centering
		\includegraphics[width=\columnwidth, clip]{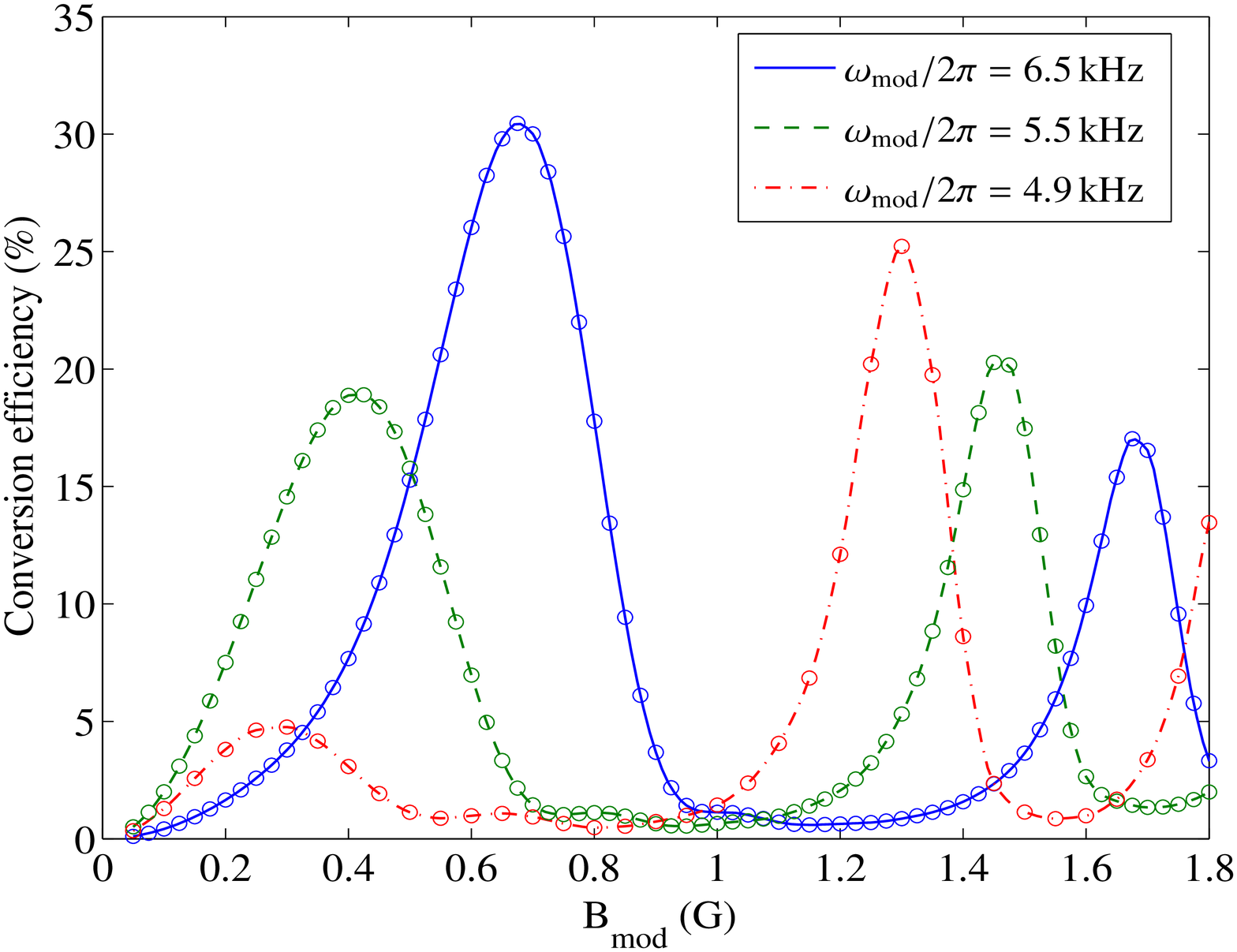}
	\caption{(Color online) Conversion from a thermal \rbf\: gas as a function of modulation amplitude for \mbox{$B_{\p{av}} = 156.352$\,G,} \mbox{$n = 10^{11}$\,cm$^{-3}$}, \mbox{$T = 20$\,nK} and \mbox{$\omega_\p{mod}/2\pi =$\:}4.9, 5.5 and 6.5\,kHz. The solid line \mbox{($\omega_{\p{mod}}/2\pi = 6.5$\,kHz)} is a thermal average of the data shown in \reffig{fig:mat_bosc}. The variation in conversion efficiency with $B_\p{mod}$ is caused by the changes in the distribution of \rhopt shown in \reffig{fig:mat_bosc}b. The revivals are caused by the regions of constructive interference, shown in \reffig{fig:mat_bosc}b, coinciding with the thermal distribution. 
	}
	\label{fig:bosc}
\end{figure}

For \mbox{$B_{\p{mod}}<0.9$\,G}, resonant growth is still observed; however, the continuum energy of the resonantly growing state is negatively shifted from that predicted by \refeq{eq:res}. As shown in \reffig{fig:mat_bosc}b, \mbox{$\rho(p, 1\,\p{ms})$} has a peak in momentum which, as $B_{\p{mod}}$ is increased, at first grows in amplitude and retains its width and position, before being shifted towards \mbox{$p = 0$}. Fully destructive interference for continuum energies up to a few kHz occurs when \mbox{$B_{\p{mod}} \approx 1.0$\,G} and so a minimum in conversion efficiency is produced, as reflected in \reffig{fig:bosc}. Beyond this modulation amplitude, there is no continuum energy for which quadratic growth of $\rho(p,t)$ is observed. \mbox{Figure~\ref{fig:mat_bosc}b} also shows two bands of constructive interference in $\rho(p,t)$. These bands have a peak energy which is also dependent on $B_\p{mod}$. Consequently, at the modulation amplitudes for which these peaks coincide with the thermal distribution, a revival of the conversion efficiency occurs. 

The maximum, minimum and revival in conversion efficiency are shown for three different modulation frequencies in Fig.~\ref{fig:bosc}. The absolute conversion efficiency at the maximum is strongly temperature dependent, as shown for $B_\p{mod} = 0.065$\,G in \reffig{fig:freq}; however, the modulation amplitude giving the maximum conversion has a weak variation with temperature above \mbox{20\,nK}. The continuum energy for which the state is resonantly coupled varies with the modulation amplitude, and so both frequency and amplitude should be matched to the temperature of the gas for maximum conversion. Of the plots shown in \reffig{fig:bosc}, for example, the best conversion is achieved for \mbox{$B_{\p{mod}} = 0.7$\,G} and \mbox{$\omega_{\p{mod}}/2\pi = 6.5 $\,kHz}. The revival in conversion efficiency for larger modulation amplitudes occurs in a peak that is narrower, and is due to the constructive interference shown in \reffig{fig:mat_bosc}b. We note that for \mbox{$B_{\p{mod}} > 1.352$\,G} the resonance position $B_0$ is being crossed during the pulse. 

\section{Conclusions}
\label{sec:conclusion}

Resonant association has been experimentally shown to be an effective technique of producing molecules~\cites{thompson05,papp06}. 
Here we have studied the dependence of the conversion efficiency on the duration, frequency and amplitude of the pulse, and the density and temperature of the gas.
We have shown that for a homogeneous gas, the continuum shapes the dynamics of the association in such a way that it is unlike a two-level system, in contrast to the case of resonant association in strongly confining optical lattices. 
The presence of other continuum states around that resonantly coupled to the Fesh\-bach molecule leads to the requirement of optimising the properties of the pulse for the gas in question.
Maximum conversion requires the amplitude and frequency of the modulation to be together optimised for the density and temperature of the gas.
Colder gases have narrower thermal distributions and so display stronger resonant behaviour. 
The width of the thermal distribution also leads to the dephasing of the oscillations in conversion efficiency observed at short times in Ref.~\cite{thompson05}. 
An increase in temperature causes a positive shift in the optimal frequency for association, but also lowers the maximum possible conversion efficiency. 

The amplitude of the modulation and mean-field shifts lead to the resonant coupling of continuum states of different energy, and thus also affect the conversion efficiency.
A higher modulation amplitude causes a less energetic continuum state to be resonantly coupled.
Beyond a certain amplitude, no resonant growth in transition probability density occurs; however, for the parameters of Ref.~\cite{thompson05} a revival in conversion efficiency is seen due to a region of constructive interference between the different continuum states.
A weak dependence of the molecular binding energy on magnetic field limits the effectiveness of resonant association, although this can sometimes be compensated for by an increase in the modulation amplitude.
The evolution of the transfer probability density from a state is primarily determined by its detuning from the resonant continuum energy.
Consequently, the pulse duration necessary for association does not vary significantly with the molecular binding energy. 
We have performed calculations for molecular binding energies ranging from $h\,\times\,$5\,kHz to $h\,\times\,$1\,MHz, and predict that resonant association can be effective over this range.

\section{Acknowledgments}
This research has been supported by the General Sir John Monash Foundation and Universities UK (T.M.H.), and the Royal Society (K.B. and T.K.). We are grateful to Sarah Thompson and Krzysztof G\'oral for interesting discussions.


\begin{thebibliography}{99}
\bibitem{donley02}
E.~A.~Donley, N.~R.~Claussen, S.~T.~Thompson, and C.~E.~Wieman, Nature (London) \textbf{417}, 529 (2002).

\bibitem{regal03}
C. A. Regal, C. Ticknor, J. L. Bohn, and D. S. Jin, Nature (London) \textbf{424}, 47 (2003).

\bibitem{review}
T. K\"ohler, K. G\'oral, and P.~S. Julienne, Rev. Mod. Phys. \textbf{78}, 1311 (2006).

\bibitem{doyle}
J.~Doyle, B.~Friedrich, R.~V.~Krems, and F.~Masnou-Seeuws, Eur. Phys. J. D \textbf{31}, 149 (2004).

\bibitem{str03}
K.~E.~Strecker, G.~B.~Partridge, and R.~G.~Hulet, Phys.~Rev.~Lett. \textbf{91}, 080406 (2003).

\bibitem{cub03}
J.~Cubizolles, T.~Bourdel, S.~J.~J.~M.~F. Kokkelmans, G.~V. Shlyapnikov, and C.~Salomon, Phys. Rev. Lett. \textbf{91}, 240401 (2003).

\bibitem{xu03}
K.~Xu, T.~Mukaiyama, J.~R.~Abo-Shaeer, J.~K.~Chin, D.~E.~Miller, and W.~Ketterle, Phys. Rev. Lett. \textbf{91}, 210402 (2003).

\bibitem{herbig03}
J.~Herbig, T.~Kraemer, M.~Mark, T.~Weber, C.~Chin, \mbox{H.-C.}~N\"agerl, and R.~Grimm, Science \textbf{301}, 1510 (2003).

\bibitem{duerr04prl}
S.~D\"urr, T.~Volz, A.~Marte, and G.~Rempe, Phys.~Rev.~Lett. \textbf{92}, 020406 (2004).

\bibitem{gre03}
M.~Greiner, C.~A.~Regal, and D.~S.~Jin, Nature (London), \textbf{426}, 537 (2003).

\bibitem{muk04}
T.~Mukaiyama, J.~R.~Abo-Shaeer, K.~Xu, J.~K. Chin, and W.~Ketterle,
Phys. Rev. Lett. \textbf{92}, 180402 (2004).

\bibitem{duerr04pra}
S. D{\"u}rr, T. Volz, and G. Rempe, Phys. Rev. A \textbf{70}, 031601
(2004).

\bibitem{volz05}
  T.~Volz, S.~D\"{u}rr, N.~Syassen, G.~Rempe, E.~van~Kempen, and S.~Kokkelmans, Phys.~Rev.~A \textbf{72}, 010704 (2005).

\bibitem{thompson05b}
S. T. Thompson, E. Hodby, and C. E. Wieman, Phys. Rev. Lett. \textbf{94}, 020401 (2005).

\bibitem{jochim03}
S.~Jochim, M.~Bartenstein, A.~Altmeyer, G.~Hendl, C.~Chin, J.~Hecker~Denschlag, and R.~Grimm, Phys.~Rev.~Lett \textbf{91}, 240402 (2003).

\bibitem{thompson05}
S. T.~Thompson, E.~Hodby, and C.~E.~Wieman, Phys.~Rev.~Lett. \textbf{95}, 190404 (2005).

\bibitem{papp06}
S. B. Papp and C. E. Wieman,  Phys.~Rev.~Lett. \textbf{95}, 180404 (2006).

\bibitem{hodby05}
E. Hodby, S. T. Thompson, C. A. Regal, M. Greiner, A. C. Wilson, D. S. Jin, E. A. Cornell, and C. E. Wieman, Phys. Rev. Lett. \textbf{94}, 120402 (2005).

\bibitem{ospelkaus06}
C. Ospelkaus, S. Ospelkaus, L. Humbert, P. Ernst, K. Sengstock, and K. Bongs, Phys. Rev. Lett. \textbf{97}, 120402 (2006).

\bibitem{bertelsen06}
J. F. Bertelsen and K. M\o lmer, Phys. Rev. A \textbf{73}, 013811 (2006).

\bibitem{child74}
M. S. Child, \textit{Molecular Collision Theory} (Academic, London, 1974).

\bibitem{moerdijk95}
A. J. Moerdijk, B. J. Verhaar, and A. Axelsson, Phys. Rev. A \textbf{51}, 4852 (1995).

\bibitem{drummond98}
P. D. Drummond, K. V. Kheruntsyan, and H. He, Phys. Rev. Lett. \textbf{81}, 3055 (1998).

\bibitem{timmermans99}
E. Timmermans, P. Tommasini, M. Hussein, and A. Kerman, Phys. Rep. \textbf{315}, 199 (1999).

\bibitem{mies00}
F. H. Mies, E. Tiesinga, and P. S. Julienne, Phys. Rev. A \textbf{61}, 022721 (2000).

\bibitem{goral04}
K. G\'oral, T. K\"ohler, S. A. Gardiner, E. Tiesinga, and P. S. Julienne, J. Phys. B \textbf{37}, 3457 (2004). 

\bibitem{koehler05}
T. K\"ohler, E. Tiesinga, and P. S. Julienne, Phys. Rev. Lett. \textbf{94}, 020402 (2005).

\bibitem{cornish00}
S. L. Cornish, N. R. Claussen, J. L. Roberts, E. A. Cornell, and C. E. Wieman, Phys. Rev. Lett. \textbf{85}, 1795 (2000).

\bibitem{taylor72}
J. R. Taylor, \textit{Scattering Theory} (Wiley, New York, 1972).

\bibitem{williams06}
J. E. Williams, N. Nygaard, and C. W. Clark, New J. Phys. \textbf{8}, 150 (2006).

\bibitem{koehler02pra}
T. K\"ohler and K. Burnett, Phys. Rev. A \textbf{65}, 033601 (2002).

\bibitem{koehler03pra}
T. K\"ohler, T. Gasenzer, and K. Burnett, Phys. Rev. A \textbf{67}, 013601 (2003).

\bibitem{koehler03prl}
T. K\"ohler, T. Gasenzer, P. S. Julienne, and K. Burnett, Phys. Rev. Lett. \textbf{91}, 230401 (2003).

\bibitem{koehler04}
T. K\"ohler, K. G\'oral, and T. Gasenzer, Phys. Rev. A \textbf{70}, 023613 (2004).

\bibitem{claussen03}
N. R. Claussen, S. J. J. M. F. Kokkelmans, S. T. Thompson, E. A. Donley, E. Hodby, and C. E. Wieman, Phys. Rev. A \textbf{67}, 060701 (2003).

\bibitem{chin04}
C. Chin, V. Vuleti\'c, A. J. Kerman, S. Chu, E. Tiesinga, P. J. Leo, and C. J. Williams, Phys. Rev. A \textbf{70}, 032701 (2004).

\bibitem{thompsonpc}
S. T. Thompson, private communication.

\end{thebibliography}
\end{document}